\documentclass[aps,twocolumn,nofootinbib]{revtex4}
\usepackage[latin1]{inputenc}
\usepackage{MnSymbol}
\usepackage{amsmath}
\usepackage{graphicx}

\begin{document}

\title{How Geometry Controls the Tearing of Adhesive Thin Films on Curved Surfaces.}

\author{Olga Kruglova}
\author{Fabian Brau}
\author{Didier Villers}
\author{Pascal Damman}
\email{pascal.damman@umons.ac.be}
\affiliation{Laboratoire Interfaces $\&$ Fluides Complexes, CIRMAP, Universit\'e de Mons, 20 Place du Parc, B-7000 Mons, Belgium}

\date{\today}

\begin{abstract}
Flaps can be detached from a thin film glued on a solid substrate by tearing and peeling. For flat substrates, it has been shown that these flaps spontaneously narrow and collapse in pointy triangular shapes. Here we show that various shapes, triangular, elliptic, acuminate or spatulate, can be observed for the tears by adjusting the curvature of the substrate. From combined experiments and theoretical models, we show that the flap morphology is governed by simple geometric rules.  
\end{abstract}

\maketitle

Cracks and fractures are commonly observed around us in various contexts~\cite{buelher10} ranging from drying mud~\cite{goeh10} to broken windows or ice floes~\cite{astr97,vell08}. The classical fracture theories, initially formulated by Griffith and Irwin~\cite{grif21,irwi57}, can reliably predict the onset of cracks motion. In contrast, no general theory is able to predict the path of a crack as it propagates. 
In some specific cases however, interesting insight has been gained about cracks trajectories. For example, thin films offer a particularly efficient set-up to study fracture propagation by limiting the crack motion to a two-dimensional manifold. In this context, the crucial role of geometry was identified in some oscillatory fracture patterns obtained when a brittle elastic thin sheet is cut by a moving object~\cite{ghat03,audo05}. The propagation of two interacting cracks in torn thin films is another example where fracture path can be understood~\cite{cohen10,fend10,bayart11}. 

\begin{figure}
\includegraphics[width=0.9\columnwidth]{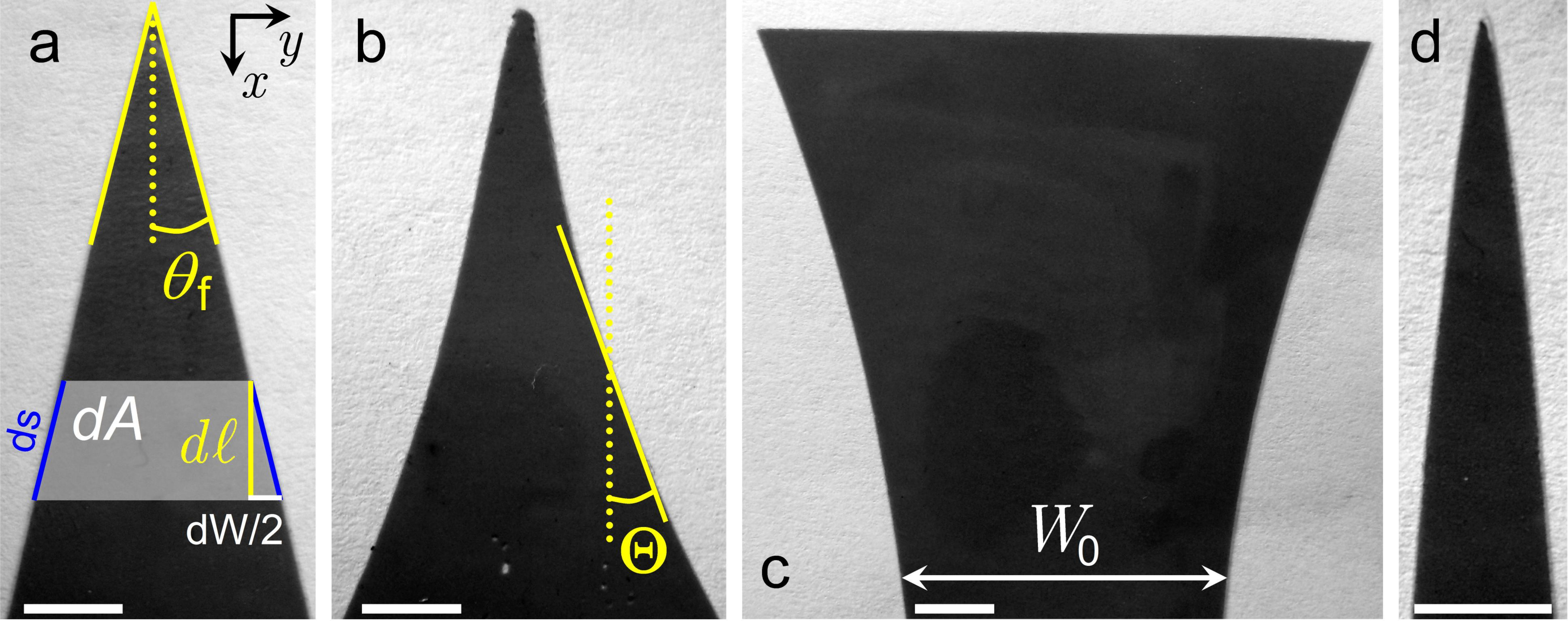}
\caption{(color online). Typical morphologies of flaps detached from adhesive tapes pulled from flat and curved substrates with a peeling angle of $90^{\circ}$. (a) Flat substrate, triangular shape. When the peeling front propagates by a infinitesimal $d\ell$ along $x$-axis, it produces two cracks of length $ds$ together with a variation of the width $dW$ and a variation of the peeling area $dA$. (b) Substrate with negative curvature, acuminate. (c, d) Substrate with positive curvature with $W_0 > W_{\text{c}}$, spatulate (c) and $W_0 < W_{\text{c}}$, elliptic (d), where $W_0$ is the initial width of the flap. Negative (positive) curvature, $W_{\text{c}}$, $\theta_{\text{f}}$ and $\Theta$ are defined in the text. Scale bars are 5 mm. The peeling front propagates upward.
}
\label{fig1}
\end{figure}

For adhesive thin films, the adhesion strength influences the crack paths. It was shown recently how thin film elasticity, adhesion and fracture act together to generate quasi-perfect triangular tears for pulled adhesive tape on flat substrates~\cite{hamm08}. Regardless of the material properties, {\it i.e.} adhesive/fracture energy, film thickness or bending modulus, the triangular morphology is extremely robust; only the tearing angle, $\theta_{\text{f}}$, is modified. This property was recently used to study the elasticity of graphene from tearing experiments~\cite{pedro10}.

In this Letter, we show through a combined experimental and theoretical study, how geometry of the substrate on which the thin film is glued gives exquisite control of the shape of the tears. With cylindrical substrates, it is possible to generate convergent and even divergent crack paths as presented in Fig.~\ref{fig1}.

\begin{figure}
\includegraphics[width=0.9\columnwidth]{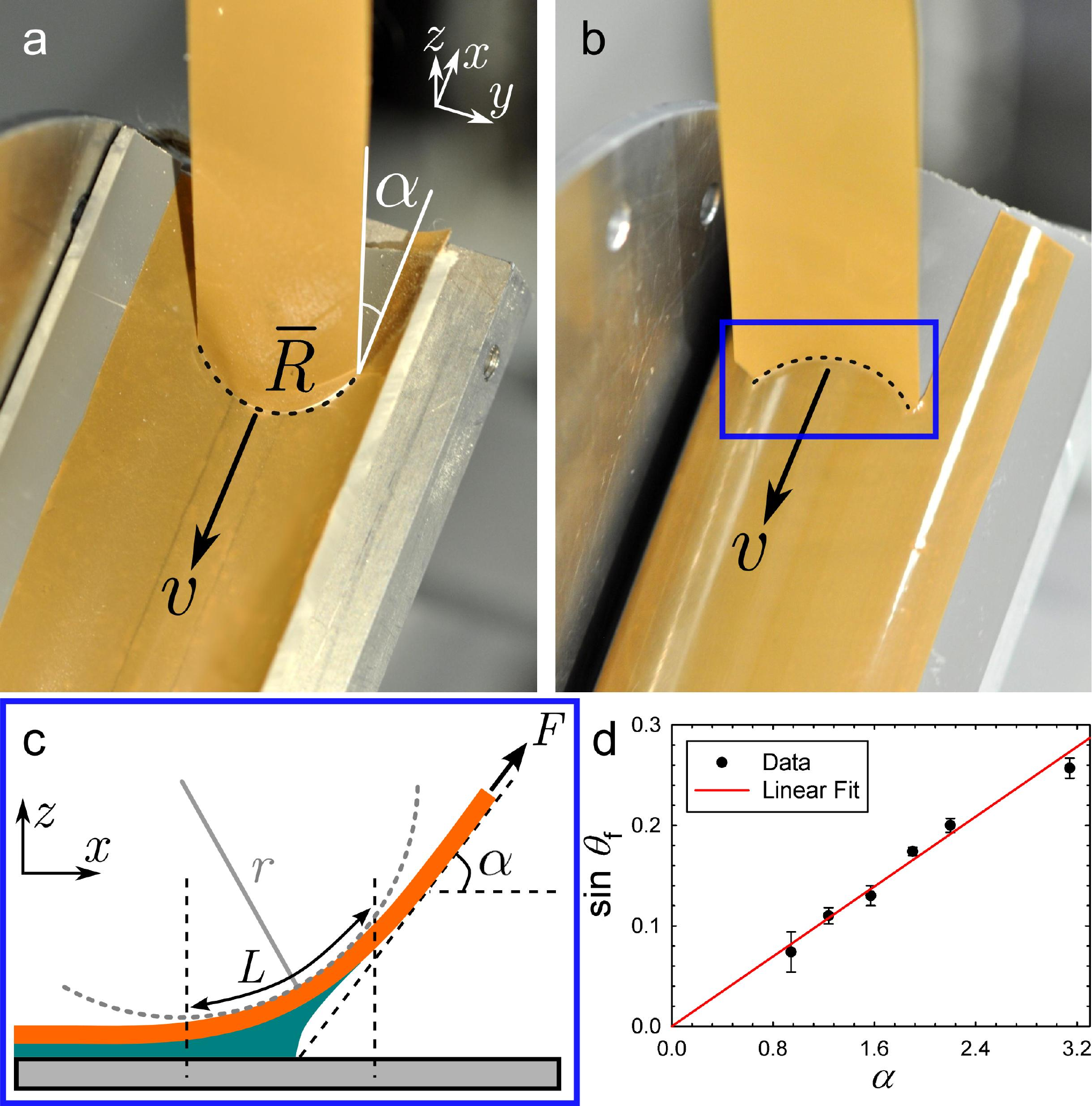}
\caption{(color online). Experimental set-up showing the peeling/tearing process for positive (a) and negative (b) curvature. The peeling front advances at a constant velocity $v$ with a constant peeling angle $\alpha$. $\bar{R}$ is the radius of curvature of the peeling front. (c) Cross-section along the cylindrical axis of the system. $r$ is the radius of curvature of the ridge connecting the flap to the substrate, $L$ is the length of this ridge and $F$ is the applied force. (d) Evolution of the tearing angle, $\theta_{\text{f}}$, for a flat substrate as a function of the peeling angle, $\alpha$. Each data point corresponds to the average of at least 5 experiments.
}
\label{fig2}
\end{figure}

In the experiments, a thin elastic sheet is adhered to various flat or curved solid substrates. Two parallel notches are cut on one edge of the adhesive film to obtain a rectangular flap. The flap is then pulled at a constant speed ($v = 1$ mm/s) and at a constant peeling angle ($54^{\circ} < \alpha < 180^{\circ}$) with a Probe Tack device (see Fig.~\ref{fig2}a, b). The final detached flaps were then digitized and further analysed (ImageJ and Plot Digitizer) to obtain the shape of the various tears (selected tears are shown in Fig.~\ref{fig1}).

{\it Flat substrates}. The shapes of tears are fully determined by the evolution of the flap width, $W$, with the peeled distance, $\ell$ during the advancing of both cracks (Fig.~\ref{fig1}). In order to explain this evolution, we use a model based on the Griffith's theory of fracture to describe the path of these tears~\cite{hamm08}. 
It should be noticed that due to the peeling geometry, the pulling force induces a deformation of the adhesive film localized in a ridge with a curvature $\kappa =1/r$ that connects the flap to the substrate, see Fig.~\ref{fig2}c.
The energy embedded into the system by pulling the flap is thus shared among the elastic energy stored in the ridge, the penalties associated to the creation of new surfaces, when cracks advance ($\gamma$, the work of fracture of the film), and the energy dissipated in the de-adhesion process ($\tau$, the adhesive energy per unit area). The variation of energy for an infinitesimal motion of the peeling front is, thus, given by 
\[
dU = dU_E + 2 \gamma t\, ds + \tau dA ,
\]
where $U_E$, $t$, $ds$ and $dA$ are the elastic energy of the flap, the film thickness, the increment in crack length, $ds = d\ell/\cos \theta_{\text{f}}$, and the variation of the peeling area $dA \simeq W d\ell$ (see Fig.~\ref{fig1}a).

Assuming that $U_E$ depends only on $W$ and $\ell$, the variation of energy needed to move the front by a distance $d\ell$ becomes ($\partial_x$ stands for $(\partial /\partial x)$),
\begin{equation}
dU = \left[-2 \partial_W U_E \tan \theta_{\text{f}} + \partial_{\ell} U_E + \frac{2 \gamma t}{\cos \theta_{\text{f}}} + \tau W \right] d\ell
\label{energy}
\end{equation}
where $dW=-2 \tan \theta_{\text{f}}\, d\ell$ since by convention a positive tearing angle leads to a decrease of the flap width. To solve this equation and find the function $\theta_{\text{f}}(\ell)$ that determines the shape of the flap, we need to compute the elastic energy stored in the ridge of curvature $\kappa$. The average curvature of this ridge is simply given by $\kappa \sim \alpha/L$, where $L$ is the length of the ridge (Fig.~\ref{fig2}c). The elastic energy is, therefore, given by 
\begin{equation}
U_E \sim BW \alpha^2/2L
\label{elastic_energy}
\end{equation}
 (where $B$ is the bending modulus of the film). 
The length of the ridge $L$ can be obtained through energy minimization, $dU/d\ell = 0$, reflecting a balance between the different energy terms of Eq.~(\ref{energy}). Considering the expression of $U_E$ given above, we obtain the relation
\begin{equation}
- \frac{B \alpha^2}{L} \tan \theta_{\text{f}} - \frac{BW\alpha^2}{2L^2} + \frac{2 \gamma t}{\cos \theta_{\text{f}}} + \tau W = 0
\label{eqforl}
\end{equation}
Assuming that the width of the flap is much larger than the length of the ridge (so that the first term of Eq.~(\ref{eqforl}) is negligible compared to the second one) and using the approximation $\cos \theta_{\text{f}}\simeq 1$ (valid for the studied range of angles), we obtain the expression for $L$, 
\[
L = \alpha \sqrt{\frac{B W}{2(2\gamma t + \tau W)}}
\]  
Inserting this expression in Eq.~(\ref{elastic_energy}) yields the final expression for the elastic energy.

The propagation of the cracks follows the angle $\theta_{\text{f}}$ which minimizes the variation of energy ({\it i.e.}, this can be considered as the maximum-energy-release-rate criterion~\cite{wu78}). The tearing angle is determined by considering that $\left(\partial_{\theta_{\text{f}}} dU\right) = 0$. 
From Eq.~(\ref{energy}), this condition becomes 
\begin{equation}
\sin \theta_{\text{f}} = \frac{\partial_W U_E}{\gamma t}
\label{sintheta}
\end{equation}
The obtained final expression for the elastic energy combined with Eq.~(\ref{sintheta}) determines the tearing angle as a function of the material properties  
\begin{equation}
\sin \theta_{\text{f}} = \frac{\alpha}{\gamma t} \sqrt{\frac{B(2 \gamma t + \tau W)}{2W}}
\label{sintheta2}
\end{equation}
Depending on the relative importance of adhesion and crack energy terms (determined by the ratio $\tau W/\gamma t$), we can distinguish two different regimes. For large flaps, adhesion dominates the evolution of the crack path ($W\gg \gamma t/\tau \simeq 1$ mm in our experiments). This regime characterized by a constant tearing angle $\theta_{\text{f}}$, is described by the relation
\begin{equation}
\sin \theta_{\text{f}} = \alpha \frac{\sqrt{2B\tau}}{2\gamma t}
\label{sintheta_f}
\end{equation}
This relation is similar to the one obtained in Ref.~\cite{hamm08}, the influence of the peeling angle $\alpha$ being added. The linear relation between $\sin \theta_{\text{f}}$ and $\alpha$ is in close agreement with the experimental data shown in Fig.~\ref{fig2}d.

In addition, a regime dominated by the fracture energy should always be observed near the tip of the flaps, {\it i.e.}, when $W$ becomes very small. This regime is characterized by $\sin \theta_{\text{f}} = \alpha \sqrt{B/\gamma t W}$. The tearing angle thus increases for decreasing flap width; this yields a shortening of the flap and a re-entrant morphology (see for instance Fig.~1 of \cite{hamm08}). Considering that $\sin \theta_{\text{f}} \simeq \tan \theta_{\text{f}} = d(W/2)/d\ell$, the shape of the tears in this regime corresponds to the power law $W\sim \ell^{2/3}$. However, it should be mentioned that this regime is rather difficult to observe when conventional adhesive tapes are used. 

\begin{figure}
\includegraphics[width=0.9\columnwidth]{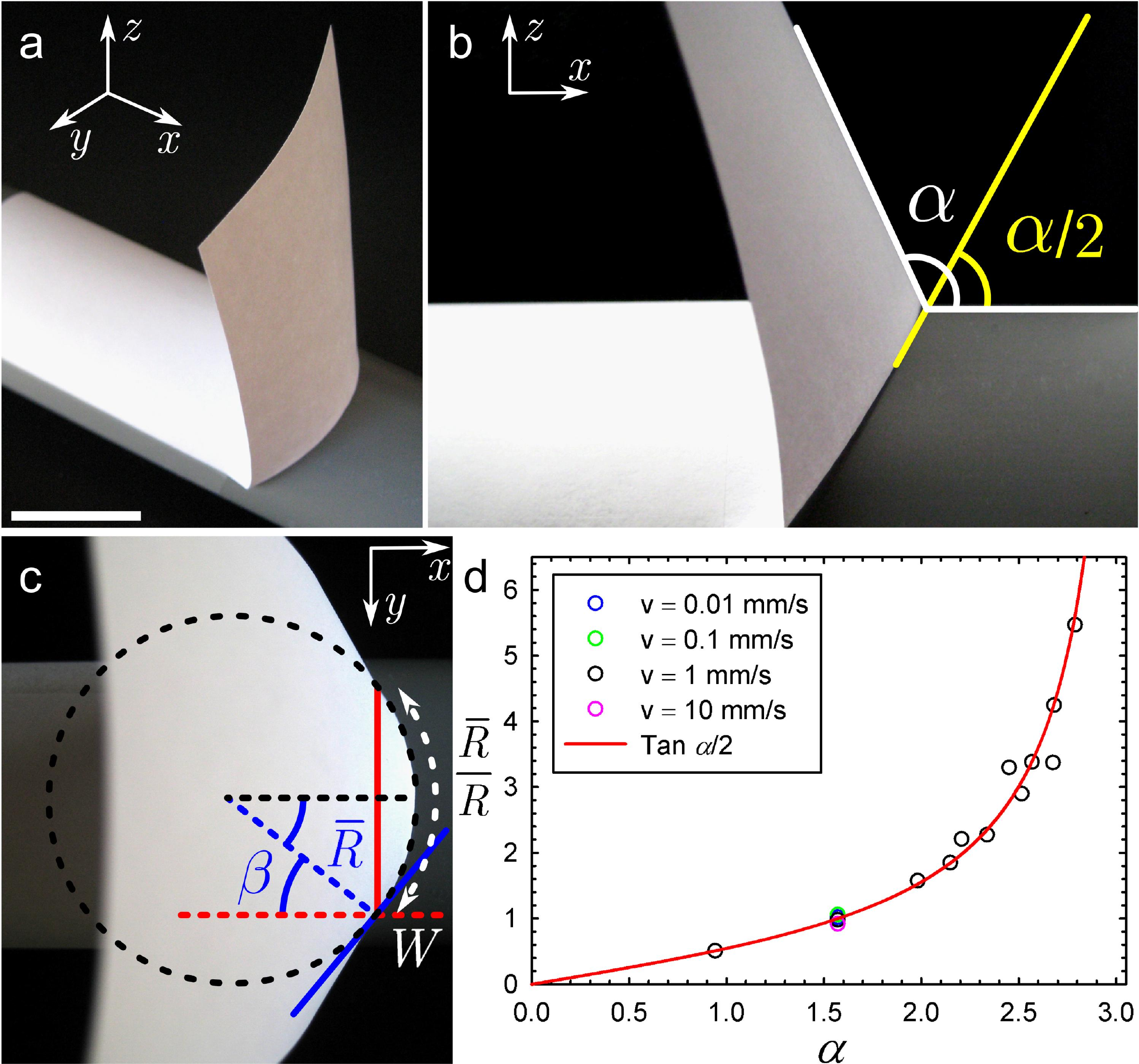}
\caption{(color online). (a) Origami model of the morphology of a thin film adhering onto a cylinder with a peeling angle $\alpha$. (b) Lateral view showing the plane containing the limiting zone of contact between the adhesive film and the cylindrical substrate and forming an angle $\alpha/2$ with $x-$axis. (c) Top view showing the curved morphology of the peeling front. The angle $\beta$ defines the rotation at the cracks of the tangent of the curved front (blue solid line) with respect to a planar peeling front (red solid line). Scale bar: 3 cm. (d) Evolution of the observed radius of curvature $\bar{R}/R$ with the peeling angle $\alpha$. 
}
\label{fig3}
\end{figure}

{\it Curved substrates}. When a thin film is glued on a curved substrate, the peeling/tearing process produces various morphologies for the flaps as depicted in Fig.~\ref{fig1}. Those shapes depend on the radius of curvature, $R$, of the substrate. 
For clarity, we define the curvature as shown in Fig.~\ref{fig2}a,b and use, in the following, the parameter $\epsilon$ equals $+1$ or $-1$ for positive or negative curvature, respectively. The main ingredient determining the global shape of the tears is the morphology of the peeling front induced by the substrate curvature. 
This front is linear for flat substrates, but, in contrast, it is characterized by a finite radius of curvature, $\bar{R}$, for cylindrical substrates as shown in Fig.~\ref{fig3}c. The direction of the curvature with respect to direction of the cracks propagation is determined by the sign of the substrate curvature. This is the main source for the breaking of symmetry which explains the occurrence of closing or opening cracks path when the peeling of the adhesive film is performed on the exterior ($\epsilon = -1$) or the interior ($\epsilon = +1$) side of the cylinder, respectively.

The relation between the radius of curvature of the peeling front and the radius of curvature of the substrate can be explained with the help of simple geometrical arguments. The limiting zone of contact between the adhesive film peeled with an angle $\alpha$ and the substrate is given by the intersection between the cylindrical substrate and a plane forming an angle $\alpha/2$ with the symmetry axis of the cylinder as shown on Fig.~\ref{fig3}b. The projection of this curve on the $(x,y)$ plane is characterized by a radius of curvature at its tip, $\bar{R}$, given by $\bar{R} = R \tan \alpha/2$, see Fig.~\ref{fig3}c. This relation corresponds to the geodesic curvature of the intersection curve~\cite{supp} and is in very good agreement with the evolution of $\bar{R}$ with the peeling angle (Fig.~\ref{fig3}d) for various values of $R$. 
The curvature of the peeling front is not significantly influenced by the speed at which the front advances, {\it i.e.} by the adhesion energy, indicating that the origin of this curved morphology is purely geometric. It is worth noting that the peeling front becomes linear for peeling angle $\alpha$ close to $180^{\circ}$ irrespective of the curvature of the substrate ($\bar{R}$ tends to infinity for $\alpha$ close to $\pi$).

In order to derive a model describing the shape of the flaps for curved substrates, we make the following assumptions: i) the tears are produced in the adhesion regime. This approximation is justified since almost everywhere the width of the flaps is large enough to neglect fracture energy ($W \gg 1$ mm) except for the extreme tip of the convergent tears; 
ii) the distortion of the peeling front induces a rotation of the path direction of the cracks by an angle $\beta$ (see Fig.~\ref{fig3}c), keeping the same value of the tearing angle $\theta_{\text{f}}$, given by Eq.~(\ref{sintheta_f}). To account for this rotation of the crack path with respect to the cylinder axis, we introduce a new angle $\Theta = \theta_{\text{f}} - \epsilon \beta$ (with $\epsilon = \pm 1$, see here above). The angle $\beta$ is determined by the distortion of the peeling front and the width of the flap and is given by $W/2\bar{R}$. This new geometry transforms the equation that determines the shape of the tears for curved substrates into  
\[
\frac{d W}{d \ell} \simeq 2\sin \Theta \simeq 2 \sin \theta_{\text{f}} - \epsilon \frac{W}{\bar{R}},
\]
with $W/\bar{R}$ as the lowest order and using as above $\cos \theta_{\text{f}} \simeq 1$. Introducing the normalization, $\tilde{W}=W/2\bar{R}\sin \theta_{\text{f}}$ and $\tilde{\ell}=\ell/\bar{R}$, we get the following ordinary differential equation (ODE)
\begin{equation}
\frac{d \tilde{W}}{d \tilde{\ell}} = 1 - \epsilon \tilde{W}
\label{equa_diff}
\end{equation}
The solution of this ODE is given by
\begin{equation}
\tilde{W} = \epsilon \left[ 1 + (\epsilon\tilde{W}_0 - 1) e^{\epsilon (\tilde{\ell}_0 - \tilde{\ell})} \right]
\label{solution}
\end{equation}
where $\tilde{W}_0=\tilde{W}(\tilde{\ell}_0)$ is an arbitrary point along the flap profile.

\begin{figure}
\includegraphics[width=0.8\columnwidth]{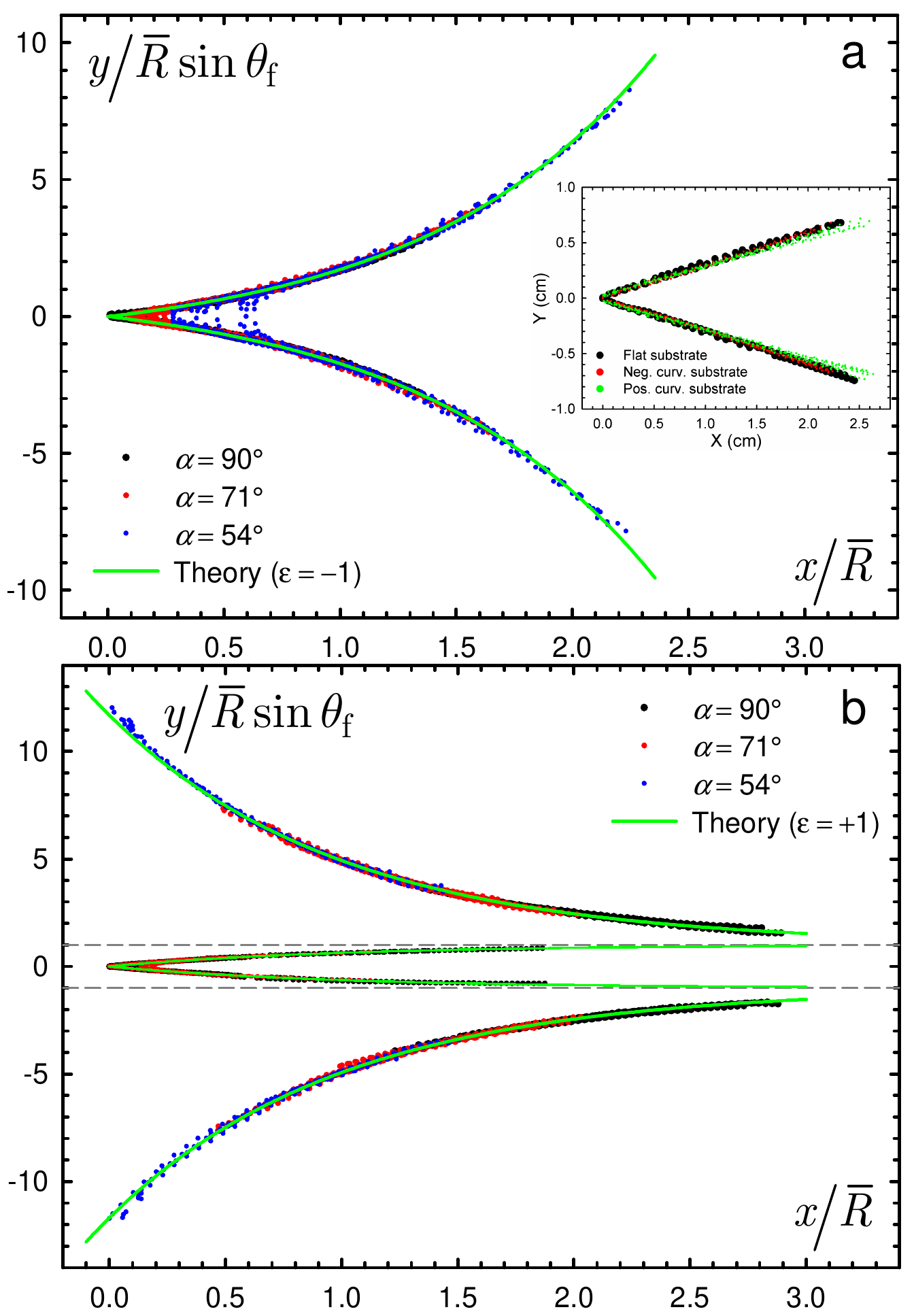}
\caption{(color online). Master curves of the tear morphologies for (a) negative curvature obtained from 45 experiments ($R =$ 54.8 , 20.2 and 16 mm) and (b) positive curvature obtained from 30 experiments ($R =$ 18.5 and 14.1 mm). The scalings used to normalize the plots are discussed in the text. The solid lines correspond to Eq.~(\ref{solution}) obtained from the model with $y=W/2$.
}
\label{fig4}
\end{figure}

Let us now discuss in detail the morphology of these solutions for the different values of $\epsilon$ and $\tilde{W}_0$. For substrates with negative curvature, we always observe closing tears with an ``acuminate leaf-like'' shape~\cite{leaf} and a sharp decrease of $\Theta$ with $\ell$ (Fig.~\ref{fig1}b). With $\epsilon=-1$ and $\tilde{W}(0)=0$ ({\it i.e.}, this point represents the ultimate tip of the tear), Eq.~(\ref{solution}) simplifies to $\tilde{W} = \left( e^{\tilde{\ell}}-1 \right)$ which captures almost perfectly the evolution of the tears shape for various samples (see the master curve in Fig.~\ref{fig4}a). 
For the positive curvature, pulling an adhesive can produce either opening (``spatulate'') or closing (``elliptic'') tears as shown in Fig.~\ref{fig1}c and d, depending on the initial value of the flap width. Interestingly, these morphologies and the transition in shape with the width are also captured by the model. For $\epsilon=+1$, the solution reads 
\begin{equation}
\tilde{W} = \left(1 +(\tilde{W}_0 -1) e^{\tilde{\ell}_0-\tilde{\ell}} \right)
\label{solution-plus}
\end{equation}
From this equation, two different regimes are indeed expected depending on the flap width.
For large initial width ($\tilde{W} > 1$ or $W > W_{\text{c}} = 2\bar{R} \sin \theta_{\text{f}}$), we observe a rapid opening of the tears contrasting with the triangular shapes previously reported. This behavior is predicted by the model since in this case, the exponential term of Eq.~(\ref{solution-plus}) is strictly positive. The agreement with the experimental data is very good, a master curve of all the results being given in Fig.~\ref{fig4}b.
For small width ($\tilde{W} <1$ or $W<W_{\text{c}}$), we recover closing tears but with an ``elliptic leaf-like'' shape instead of an acuminate one. This behavior is also predicted by our model since in this case, the exponential term of Eq.~(\ref{solution-plus}) is strictly negative. Again, the model based on geometry arguments captures the variety of the tear morphology observed for curved substrates. The transition between opening and closing crack paths predicted by the model is also experimentally observed and perfectly reproduced by the theory (Fig.~\ref{fig4}b).
It should be emphasized that, for positively curved substrates and a given set of experimental conditions ($R$, $\alpha$, $\gamma$, $\tau$, and $t$) there is a single value of the width $W_{\text{c}} = 2\bar{R} \sin \theta_{\text{f}}$ that produces tears with a constant width and parallel cracks. This value obviously corresponds to an unstable state lying between both opening and closing domains. 

As shown here above, the peeling front is linear for peeling angle close to $\pi$ regardless of the curvature of the substrate. As expected from the proposed model, the tears adopt a triangular morphology for various curved substrates when the flap is pulled at $180^{\circ}$, see the inset of Fig.~\ref{fig4}a.

In conclusion, we have demonstrated for flaps detached from films glued to simple circular cylinders that a constant mean curvature strongly affects the crack path and thus the shape of the tears. Moreover, the proposed model can be readily extended to any developable surface, {\it i.e.} conical or cylindrical. Indeed, Eq.~(\ref{equa_diff}) describes the local behavior of the crack propagation and, therefore, a spatial variation of the surface curvature can be easily introduced.  
A possible extension of this work could be the study of the influence of Gaussian curvature on crack path and tear morphology for surfaces such as sphere, hyperboloid.

This work was partially supported by the Belgian National Funds for Scientific Research (FNRS), the Government of the Region of Wallonia (REMANOS Research Programs), the European Science Foundation (Eurocores FANAS, EBIOADI).

\clearpage

{\bf {\it Supplemental Material for} ``How Geometry Controls the Tearing of Adhesive Thin Films on Curved Surfaces''}

\section{Peeling front curvature}

\begin{figure*}
\includegraphics[width=0.8\textwidth]{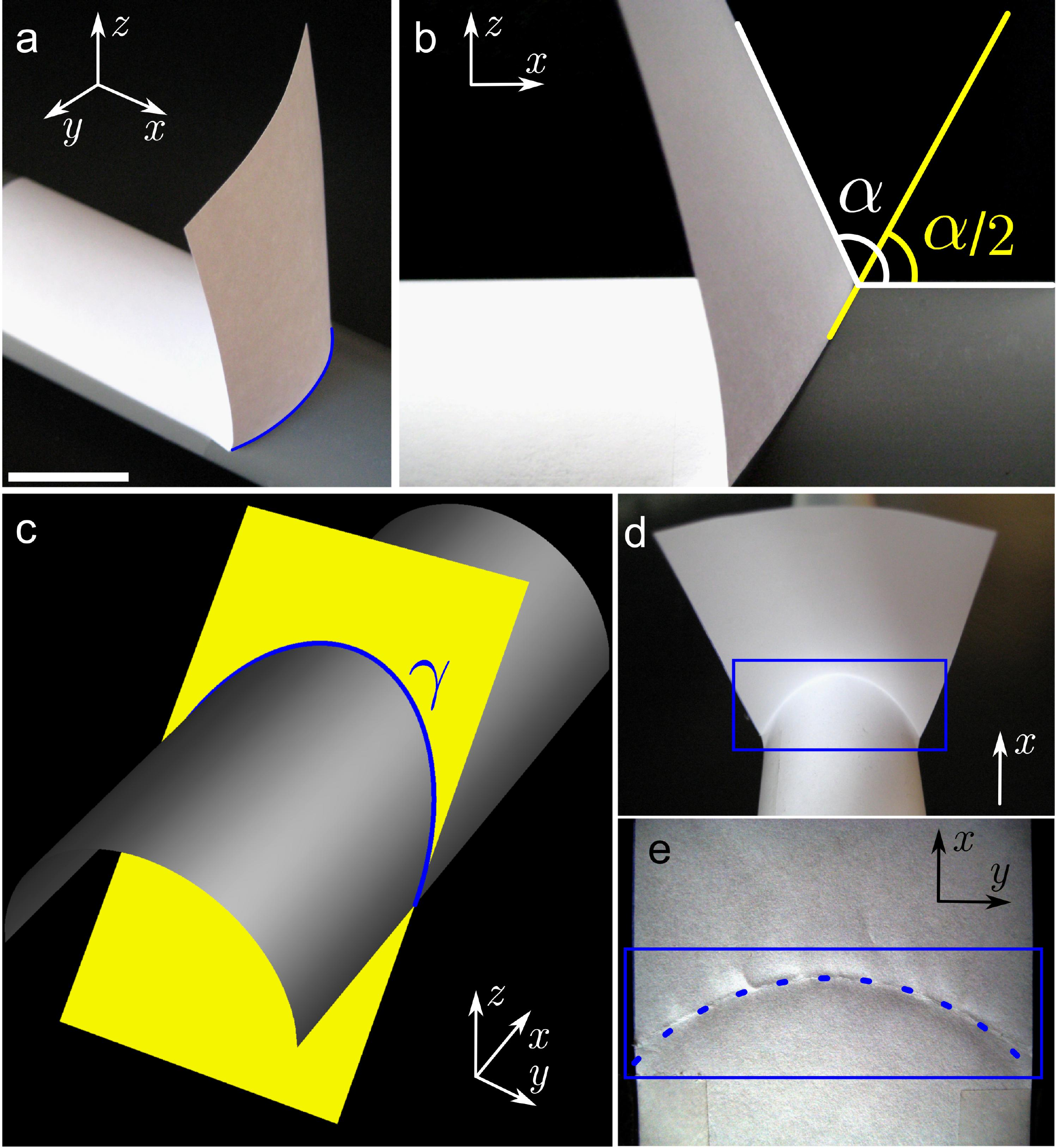}
\caption{(a) Origami model of the morphology of a thin film adhering onto a cylinder with a peeling angle $\alpha$.  Scale bar: 3 cm. (b) Lateral view showing the plane containing the limiting zone of contact between the adhesive film and the cylindrical substrate and forming an angle $\alpha/2$ with $x-$axis.  (c) Intersection curve $\gamma$  between a cylinder and a tilted plane. (d) View of the wrapped paper sheet around a cylinder (total curvature). (e) View of the flattened paper sheet (geodesic curvature), the creasing of the paper sheet reveals the intersection curve which is compared to an arc of circle (blue dotted line).} 
\label{fig01}
\end{figure*}

In Fig.~\ref{fig01}a below (see also Fig.~3 of the main text), a sheet of paper is wrapped around a cylinder to model the shape adopted by the thin adhesive film during our peeling/tearing experiments. The blue solid line indicates the peeling front whose shape is identified to a curve $\gamma$ obtained by the intersection between a cylinder and a plane forming an angle $\alpha/2$ with the $x$-axis (see Fig.~\ref{fig01}b,c). When the sheet is unrolled and flattened, the creasing of the paper sheet reveals the curvature of the peeling front relevant for our study (see Fig.~\ref{fig01}d,e). This procedure allows to distinguish the part of the curvature of the curve $\gamma$ that is due to the cylinder and the one that comes from the peeling front itself. In mathematical terms, this means that we are dealing with the {\it geodesic curvature}, $\kappa_g = (R_g)^{-1}$, of the curve $\gamma$ and not with the total curvature. As illustrated in Fig.~\ref{fig01}e, this geodesic curvature varies slowly along the peeling front and is approximated in our study by the curvature, $\bar{R}^{-1}$, at the tip of the front (at $y=0$). The deviation between the peeling front and an arc of circle is also shown. The accuracy of this approximation improves as the ratio between the width of the front, $W$, and the radius of the cylinder, $R$, decreases.

\subsection{Equation for the intersection curve $\gamma$}

The equation of the cylinder is given by
\begin{equation}
\label{cyl}
z^2+y^2=R^2,
\end{equation}
with $-\infty<x<\infty$. The equation of the plane reads
\begin{equation}
\label{plane}
z=\tan(\alpha/2) x,
\end{equation}
with $-\infty<y<\infty$. Consequently, the parametric equations for the intersecting curve $\gamma$ is found to be
\begin{equation}
\label{eq-curve}
\vec{\gamma}(t)=\left(t,\pm \sqrt{R^2-\tan^2(\alpha/2) t^2},\tan(\alpha/2) t\right),
\end{equation}
with $-R/\tan(\alpha/2)\le t \le R/\tan(\alpha/2)$. 

\subsection{Expression of $\bar{R}$: simple derivation}

Since we are only interested by the geodesic curvature at the {\it tip} of the curve $\gamma$, there are two equivalent ways to compute this quantity. The first one is obviously to directly compute the geodesic curvature along the curve $\gamma$ by using the general definition of this quantity, as detailed in the next section. The second equivalent one is to compute the curvature at the tip of the curve $\gamma_{xy}$ obtained by projecting the curve $\gamma$ in the plane $(x,y)$, as mentioned in the main text.

From Eqs.~(\ref{cyl}) and (\ref{plane}), this curve, $\gamma_{xy}$, is an ellipse whose equation reads
\begin{equation}
y^2+\tan^2(\alpha/2) x^2=R^2.
\end{equation}
The radius of curvature, $\bar{R}$, at the tip $y=0$ is thus given by
\begin{equation}
\label{rbar}
\bar{R} = R \tan(\alpha/2).
\end{equation}
This corresponds to the expression used in the main text and tested experimentally (see Fig.~3d of the main text).

\subsection{Expression of $\bar{R}$: geodesic curvature}

An alternative procedure to obtain the result (\ref{rbar}) is to compute the geodesic curvature defined by the relation~\cite{slob02} 
\begin{equation}
\kappa_g(t)=\frac{\vec{\gamma}''(t)\cdot(\vec{n}(t)\times \vec{\gamma}'(t))}{||\vec{\gamma}'(t)||^3}
\end{equation}
where prime denotes the derivative with respect to the parameter $t$, $||\vec{x} ||$ is the norm of $\vec{x}$ and $\vec{n}$ is the unit normal to the cylinder. The radius of curvature $\bar{R}$ is given by
\begin{equation}
\bar{R}=(|\kappa_g|)^{-1},
\end{equation}
where $\kappa_g$ is evaluated at $t=R/\tan(\alpha/2)$ which correspond to $y=0$ from Eq.~(\ref{eq-curve}). 

The unit normal to the cylinder is computed from the gradient of $G(x,y,z) = z^2+y^2-R^2$ as follow
\begin{equation}
\vec{n}(x,y,z)=\frac{\vec{\nabla}G}{||\vec{\nabla}G||}.
\end{equation}
The expression of this normal vector along the curve $\gamma$ is obtained by replacing $x,y$ and $z$ in terms of $t$ using the parametrization (\ref{eq-curve}):
\begin{equation}
\vec{n}(t)=\left(0, \sqrt{1-\tan^2(\alpha/2) (t/R)^2},(t/R)\tan(\alpha/2)\right).
\end{equation}
The rest of computation is done without difficulty and we obtain
\begin{eqnarray}
R_g(t)&=&(\kappa_g(t))^{-1} \\
      &=&R\frac{\left(1+\tan^2(\alpha/2)-(t/R)^2\tan^2(\alpha/2) \right)^{3/2}}{(t/R)\tan^3(\alpha/2)}  \nonumber
\end{eqnarray} 
Finally, the geodesic curvature at the tip of the curve $\gamma$ is given by
\begin{equation}
\bar{R}=R_g(R/\tan(\alpha/2))=R\tan(\alpha/2)
\end{equation}


\begin{thebibliography}{99}
\bibitem{leaf} For clarity, we choose the chart of leaf morphology to describe the global shape of the tearing flaps.
\end{thebibliography}
\end{document}